\documentclass[10pt,aps,showpacs,superscriptaddress,twocolumn,longbibliography,nofootinbib]{revtex4-2}
\usepackage{xcolor}
\usepackage{amsmath}
\usepackage{mathtools}
\usepackage{amssymb}
\usepackage{amsfonts}
\usepackage{amsthm}
\usepackage{mathrsfs}
\usepackage{footnote}
\usepackage{graphicx} 
\usepackage{stmaryrd}
\usepackage{comment}
\usepackage[colorlinks=true,linkcolor=blue,citecolor=blue]{hyperref}
\newtheorem{thm}{Theorem}
\newtheorem{lem}{Lemma}

\newcommand{\ket}[1]{| #1 \rangle}

\newcommand{\bra}[1]{\langle #1 |}

\newcommand{\ketbra}[2]{\left|#1\right\rangle\!\!\left\langle #2\right|}

\def\eps{\varepsilon}
\def\O{\mathcal{O}}

\def\id{\mathbb{I}}
\newcommand*{\tr}{\mathrm{Tr}}

\begin{document}

\title{Quantum Energy Teleportation versus Information Teleportation}
\author{Jinzhao Wang}\email{jinzhao@stanford.edu}
\author{Shunyu Yao}\email{shunyu.yao.physics@gmail.com}
\affiliation{Stanford Institute for Theoretical Physics, Stanford University, Stanford, CA 94305}
\begin{abstract}
    Quantum energy teleportation (QET) is the phenomenon in which locally inaccessible energy is activated as extractable work through collaborative local operations and classical communication (LOCC) with an entangled partner. It closely resembles the more well-known quantum information teleportation (QIT) where quantum information can be sent through an entangled pair with LOCC. It is tempting to ask how QET is related to QIT. Here we report a first study of this connection. Despite the apparent similarity, we show that these two phenomena are not only distinct but moreover are mutually competitive. We show a perturbative trade-off relation between their performance in a thermal entangled chaotic many-body system, in which both QET and QIT are simultaneously implemented through a traversable wormhole in an emergent spacetime. Motivated by this example, we study a generic setup of two entangled qudits and prove a universal non-perturbative trade-off bound. It shows that for any teleportation protocol, the overall performance of QET and QIT together is constrained by the entanglement resource. We discuss some explanations of our results.
\end{abstract}

\maketitle

\emph{Introduction.}---Quantum (information) teleportation (QIT) is a protocol in which the quantum entanglement shared between two distant agents can be turned into quantum communication with the help of local operations and classical communication (LOCC)~\cite{bennett1993teleporting,bouwmeester1997experimental,boschi1998experimental}. Importantly, the quantum information cannot be sent with either entanglement or LOCC alone. The discovery of quantum teleportation marks a pivotal moment in unraveling the mysteries of entanglement and leveraging its potential in quantum technology. 

Could one teleport something other than information when combining the resource of entanglement with LOCC? Quantum energy teleportation (QET) is the next of kin~\cite{hotta2008protocol,hotta2008quantum}. Analogously to QIT, QET allows Alice to teleport energy to Bob using entanglement and LOCC. QET was originally devised in quantum field theory as a means to create negative energy density. It is later generalized to other contexts like spin chains~\cite{hotta2009quantum} or entangled qubits~\cite{hotta2010energy}. Recently, it has been experimentally demonstrated both in the lab~\cite{rodriguez2023experimental} and on a quantum chip~\cite{ikeda2023demonstration}.

Here is how QET works. Suppose Bob would like to locally extract some energy/work from the system he has. Without accessing auxiliary systems like a thermal bath, Bob cannot extract energy if his system is in a \emph{passive} state, such as the thermal state or the ground state. Mathematically, passivity means whatever unitary Bob applies to his system makes the energy higher, so by energy conservation, Bob is injecting rather than extracting energy. However, if Bob's system is entangled with a part shared by his partner Alice, Bob could use Alice's help by asking her to make some measurements on her system and communicating the results to Bob. Bob then applies conditional operations based on Alice's results. It turns out that when acted in accordance, such a strategy could allow Bob to inject negative energy into the system and hence equivalently to energy extraction.

One immediate question that comes to our mind is how is QET related to QIT? Could we achieve both goals with one protocol? Naively, we thought QET and QIT should go hand in hand simply because they work analogously. 

To answer this question, we examine a setup where QIT and QET are simultaneously implemented with a single protocol. It concerns two entangled strongly coupled many-body quantum systems whose complicated dynamics can be more easily understood with the help of an emergent spacetime~\cite{KitaevTalks,maldacena2016conformal,maldacena2016remarks}. For a thermal field double state entangling two such systems, a special family of information teleportation protocols can be described through the physics of a traversable wormhole in the emergent spacetime~\cite{gao2017traversable,maldacena2017diving}. This has inspired a series of works exploring the deep connection between teleportation and wormholes~\cite{susskind2018teleportation,maldacena2018eternal,gao2021traversable,brown2023quantum,nezami2023quantum,schuster2022many,jafferis2022traversable,zhou2024size,liu2024fidelity}. We observe that the same protocol also simultaneously teleports energy. Interestingly,  we find QET and QIT are competitive with each other in this example. Their exclusivity is manifest in the perturbative trade-off relation that we demonstrate. 

Motivated by this observation, we look for \emph{universal} non-perturbative trade-off relations between QET and QIT. To this end, we resort to the standard finite-dimensional setup that consists of two arbitrarily entangled systems with some arbitrary non-interacting Hamiltonian. We prove a \emph{universal non-perturbative trade-off bound} for both the thermal-entangled states (Theorem~\ref{thm:1}) and arbitrary entangled states (Theorem~\ref{thm:2}), showing that the overall performance of QET and QIT altogether is upper-bounded by the amount of shared entanglement. The bounds are universal and non-perturbative in the sense that they apply to any pure entangled resource states, any measurement schemes of Alice and any conditional unitaries of Bob. The two theorems consitute the main results of this paper.

\emph{A motivating example: Traversable wormholes.}---We find that a setup that naturally incorporates both teleportations is the \emph{traversable wormhole} in holographic systems~\cite{gao2017traversable,maldacena2017diving}. It was discovered to be a convenient gravitational description of QIT. We use a concrete example of such systems in AdS$_2$/CFT$_1$ correspondence~\cite{maldacena2017diving} to explain the result. The bulk theory is described by Jackiw–Teitelboim (JT) gravity. The boundary theory is described Schwarzian quantum mechanics\footnote{Technically, Schwarzian quantum mechanics has a continuous energy spectrum. However, the Schwarzian describes the low energy dynamics of the Sachdev-Ye-Kitaev (SYK) model at the large $N$ limit~\cite{KitaevTalks}, so we are allowed to think of the boundary dual of JT gravity as a normal quantum mechanical system like the SYK model with $N$ fermions.}, such as the Sachdev-Ye-Kitaev (SYK) model~\cite{KitaevTalks}.

We consider a two-sided black hole with inverse temperature $\beta$, dual to the thermal field double state~\cite{maldacena2003eternal},
\begin{equation}\label{eq:tfd}
    \ket{\tau_\beta}_{AB} := Z^{-\frac12}\sum_x e^{-\beta E_x/2}\ket{E_x}_A\ket{E_x}_B\ ,
\end{equation}
where the ${E_x, \ket{E_x}}_x$'s are the eigenvalues and eigenvectors of the Hamiltonian $H_A=H_B$, and $Z:=\tr\,e^{-\beta H_B}$ is the partition function normalizing the state.

\begin{figure}
    \centering
    \includegraphics[width=0.4\textwidth]{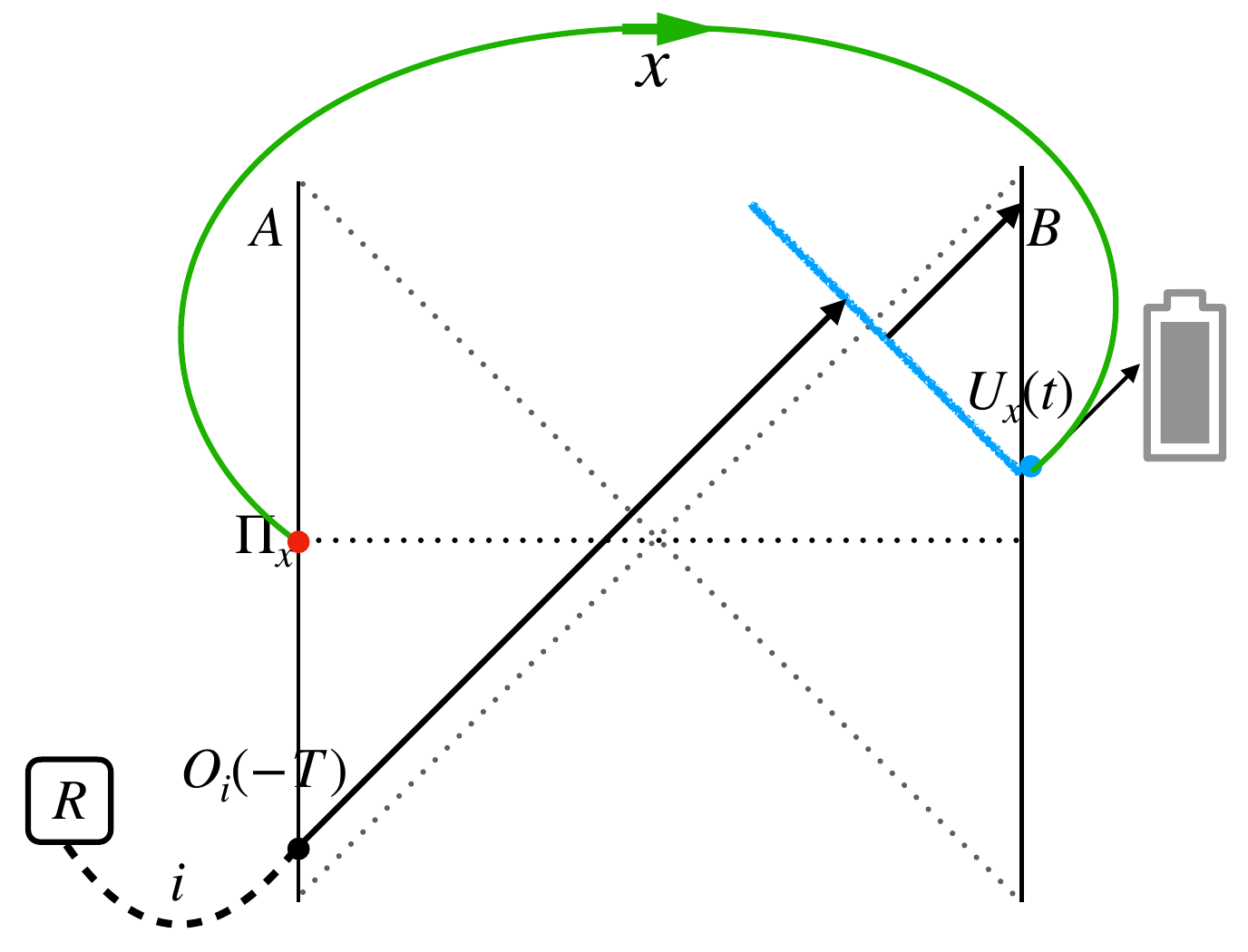}
    \caption{\textbf{Traversable Wormhole Teleportations.} The effect caused by the double trace deformation $e^{igV_{AB}}$ on Bob is equivalent to the LOCC consisting of Alice's measurement $\Pi_x$ at time zero (red dot), classical communication (green arrowed line) and the Bob's unitary $U_x(t)$ at time $t$. The message referenced by $R$ inserted by Alice at time $-T$ is sent to Bob via the LOCC. In the bulk, the dual message particle passes through the wormhole that is made traversable by the negative energy shock wave created via $U_x(t)$. The protocol $(\Pi_x,U_x)$ simultaneously achieves both QET and QIT.} 
    \label{fig:TW}
\end{figure}

The wormhole in Fig.~\ref{fig:TW} is a priori not traversable, as shown by the diagonal dotted causal wedges in the Penrose diagram. Any two points between the left and right causal wedges are space-like separated. On the boundary, this indicates an entangled state alone can not be used for communication. However, seminal works~\cite{gao2017traversable,maldacena2017diving} point out that the wormhole can be made traversable by introducing a double trace deformation on two sides $e^{igV_{AB}}$, where $V_{AB}:=\frac1K\sum_{i=1}^K O_i^A O_i^B$. This interaction couples $K$ pairs of simple operators $O_i$ on $\O(1)$ of fermions and $g$ controls the strength. The deformation with $g>0$ creates an ingoing shock wave with negative null energy that moves the black hole horizons inward, rendering a traversable wormhole\footnote{For this description to be accurate, one technically chooses $K$ to be large $K\sim\O(N)$ so as to have a sharp bulk dual as a classical shock-wave geometry.}. Namely, a particle carrying information thrown in the past from Alice's side can now make its way to Bob's side. 

As far as Bob's system is concerned, one can view the effect of the double trace deformation $e^{igV_{AB}}$ on Bob's system, as an LOCC protocol $\prod_x\Pi_x^A\otimes U^B_x $, in which Alice measures her share of the entanglement pair and Bob implements a conditional unitary on his share upon receiving the measurement result from Alice after a time delay of~$t$ (cf. Fig.~\ref{fig:TW} and Appendix~\ref{app:LOCC} for details). Usually, $e^{igV_{AB}}$ is acted on the time-symmetric slice so $t=0$, but we consider acting it asymmetrically to fit the operational interpretation as an LOCC, so a delay of $t>0$ is inevitable. In this protocol, Alice can transmit information by inserting a simple operator with a species index at around the scrambling time $T$ before the measurement. The information will be transmitted to Bob's side after Bob's unitary decoding. After a scrambling time, the information will re-focus onto simple operators. (cf.~\cite{gao2021traversable} for details in the context of the SYK model.)

The first observation we make is that the traversable wormhole protocol actually also implements quantum energy teleportation. Without Alice's communication, Bob cannot extract energy by doing a unitary, because Bob's reduced density matrix is a thermal density matrix. With Alice's measurement result $x$, Bob applies the conditional unitary decoder $U_x$ after some communication delay $t$. By energy conservation, Bob then gains energy that amounts to the negative energy that is injected into his system via the conditional unitary. In Appendix~\ref{app:TW}, we calculate this teleported energy,
\begin{equation}\label{Energychange}
   E= g\Delta\left(\frac{2\pi}{\beta}\right)^{2\Delta+1} \frac{\tanh(\frac{\pi}{\beta}t)}{\cosh^{2\Delta}(\frac{\pi}{\beta}t)}\  ,
\end{equation}
where $\Delta$ is the conformal weight of the deformation operators $\{O_i\}$ in $V_{AB}$. 

Now we can achieve QIT and QET simultaneously with the same protocol. However, sending information degrades the QET. This is manifest from the bulk perspective. The Bekenstein bound implies that the particle carrying the information in QIT, must also carry positive energy~\cite{bekenstein1981universal}. Hence, sending the information via a particle lowers the amount of extractable energy by an amount of $\delta E$. This hints at a trade-off relation between QET and QIT for the traversable wormhole protocol. 

We can make this trade-off more precise by quantifying the teleported information. 
We use $k$ different particle species indexed by $x\in [k]$ to carry the message. To calculate how much information can be teleported, we consider sending a maximally mixed message. By linearity, the state that Bob receives is $\rho=\frac1k\sum_{x=1}^k\rho_x$, where $\rho_x$ is the protocol output for the message $x$. The entropy of the message is measured by $S:=S(B)_{\rho}-S(B)_{\rho_0}$, where $\rho_0$ is Bob's density matrix after the double trace deformation but without the message particle insertion\footnote{One could use other measures such as the coherent (Holevo) information to quantify the quantum (classical) information capacity. They all behave very similarly to the subtracted entropy w.r.t. varying k~\cite{bousso2017universal,hayden2023exactly}. We therefore use the simplest figure of merit as a proof of principle.}. 

We can calculate each term in $S$ using the (quantum) Ryu-Takayanagi formula~\cite{ryu2006holographic,faulkner2013quantum}. The state $\rho_0$ describes a bulk without matter and its entropy is given by $S(B)_{\rho_0}=A/4G_N (=\beta\langle H_B\rangle_{\rho_0})$, where $A$ is the area of the shifted bifurcation surface after the deformation, and it follows from the thermality of $\rho_0$\footnote{One can show this using the generalized gravitational entropy~\cite{lewkowycz2013generalized,faulkner2013quantum}}. On the other hand, $S(B)_{\rho} = A/4G_N + S_\mathrm{matter}$\footnote{Here the area term will be the same as for $\rho_0$ under the approximation of the messenger particle being light.}, where $S_\mathrm{matter}$ is the quantum field theoretic entropy of the message particle. We know that $S_\mathrm{matter}$ of a uniform mixture of $k$ particle species grows like $\log k$ until it saturates at the Bekenstein bound set by $\beta E$, and $E:=\langle H_B\rangle_\rho -\langle H_B\rangle_{\rho_0}$.
Hence, $S$ increases with $\log k$, but eventually caps off due to the Bekenstein bound~\cite{bekenstein1981universal,marolf2004notes,marolf2005few,casini2008relative}. The maximal $S$ thus measures the capacity of QIT\footnote{This calculation is explicitly done using the replica trick in the upcoming paper by Lu-Yang-Zheng~\cite{yang2024}.}.

Therefore, the loss in QET is totally compensated by the gain in QIT\footnote{As always in thermodynamics, the (inverse) temperature $\beta$ here is the conversion factor that is necessary for comparing the distinct resources of information and energy.}. In other words, upon varying the relevant parameters of the protocol, such as the conformal weight the message particle, the total variation of the combined QET+QIT figure of merit $S+\beta E$ is zero,
\begin{equation}\label{eq:result0}
    \delta(S+\beta E) = \delta S + \beta\delta E = 0\ .
\end{equation}
The bulk interpretation of this perturbative trade-off is the saturation of Bekenstein bound\footnote{This is also reminiscent of the first law of entanglement~\cite{blanco2013relative}, that is derived by varying the relative entropy around zero. According to Casini~\cite{casini2008relative}, the zero relative entropy between Bob's reduced state and the thermal background is precisely the point of the Bekenstein bound saturation.}. Given the energy deficit $\delta E$, the maximal entropy it can carry is equal to the modular energy of Bob's causal wedge, which is $\beta\delta E$ in terms of the boundary ADM energy. 

There are some drawbacks of the trade-off~\eqref{eq:result0}. It is only perturbative because we cannot vary $m$ too much, as it causes significant backreaction rendering our semiclassical analysis inaccurate. Moreover, it only applies to a specific family of LOCC schemes that admit the bulk description as traversable wormholes. In seeking of a non-perturbative trade-off relation that applies to all LOCC protocols, we resort to a generic finite-dimensional setup.

\emph{Main results: Trade-off between QET and QIT.}---We will now prove two trade-off bounds for the thermal field double state and arbitrary pure entangled states respectively. We first consider a finite-dimensional version of~\eqref{eq:tfd}, $\ket{\tau_\beta}_{AB}$, between Alice and Bob at the inverse temperature $\beta$. Let Bob's Hamiltonian be $H$\footnote{Since we do not care about the energy change of Alice's system, it is not necessary to specify a Hamiltonian for Alice and we can also let the $\ket{E_x}_A$ to be any basis.}. 

\begin{figure}
    \centering
    \includegraphics[width=0.5\textwidth]{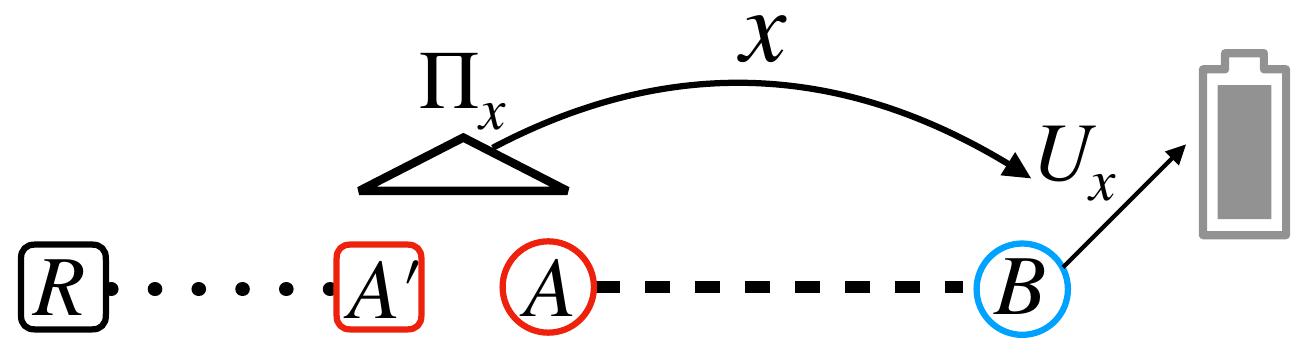}
    \caption{\textbf{QET and QIT in finite dimensions.} The dashed line represents the entanglement resource shared by Alice and Bob. The dotted line represents the maximal entanglement to be swapped with Bob. Alice measures $AA'$ with $\Pi_x$, $x$ is the result sent to Bob and Bob applies $U_x$. A teleportation protocol $(\Pi_x,U_x)$ simultaneously achieves both QET and QIT.}
    \label{fig:qudit}
\end{figure}

The state $\ket{\tau_\beta}_{AB}$, or rather just Bob's reduced density matrix $\tau_\beta=e^{-\beta H}/Z$, is passive as no local unitaries on $B$ can extract energy from it. The state $\ket{\tau_\beta}_{AB}$ shall be our resource state for both QET and QIT, and the latter also demands an unknown target state to teleport. A standard procedure is to consider teleporting the $A'$ part of a maximally entangled pair $\ket{\Phi}_{A'R}$ where $R$ is a reference, also known as \emph{entanglement swapping} (cf. Fig.~\ref{fig:qudit})~\cite{yurke1992einstein,zukowski1993event,pan1998experimental,horodecki2009quantum}. Success in preserving the entanglement in $\ket{\Phi}$ is equivalent to the ability to teleport any unknown quantum state~\cite{horodecki1999general,nielsen2002simple}. A common figure of merit for a QIT protocol is the entanglement fidelity, which is simply the fidelity between $\ket{\Phi}_{A'R}$ and the output state on $RB$. However, fidelity itself is not an entropic measure and we find it easier to work with entropic measures in formulating a trade-off between energy and information. We consider the same figure of merit $S+\beta E$ as in~\eqref{eq:result0}, and aim to prove a universal converse bound on it in terms of the amount of entanglement resource in $\ket{\tau_\beta}$.

We focus on teleportation protocols with rank-one projective measurements on $AA'$, as general measurements with higher-rank Kraus operators have coarser resolution. Since we are interested in the trade-off relation as constrained by the entanglement resource, it is sufficient to consider measurements with resolution as sharp as possible\footnote{One could also ask how the resource of classical communication constrains the teleportations. Then it is also of interest to consider general measurements whose outcome consumes less communication resource to be sent.}. For completeness, we discuss the case of general measurement instruments in Appendix~\ref{app:mmts}. With rank-one measurement operators, the output states on $RB$ are pure so it is appropriate to use the entanglement entropy as a figure of merit for QIT.  We prove the following result\footnote{There are some variants of the trade-off relation~\eqref{eq:result1} that we discuss in Appendix~\ref{app:variant}.}.

\begin{thm}\label{thm:1}
Given the resource state $\ket{\tau_\beta}_{AB}$ as in~\eqref{eq:tfd} and a Bell state $\ket{\Phi}_{A'R}$, any teleportation protocol, consisting of rank-one projective measurement $\{\Pi_x\}$ on $AA'$ followed by any conditional unitaries $\{U_x\}$ on $B$, satisfies
\begin{equation}\label{eq:result1}
    S(B)+\beta E \le S(B)_{\tau_\beta}\ ,
\end{equation}
where $S(B):= \sum_x p_x S(B)_{\rho_x}$, $E:= \sum_x p_x E_x$, $\rho_x$ is the output state upon obtaining outcome $x$ and $E_x$ is the extracted energy from it. 
\end{thm}

\begin{figure}
    \centering
    \includegraphics[width=0.4\textwidth]{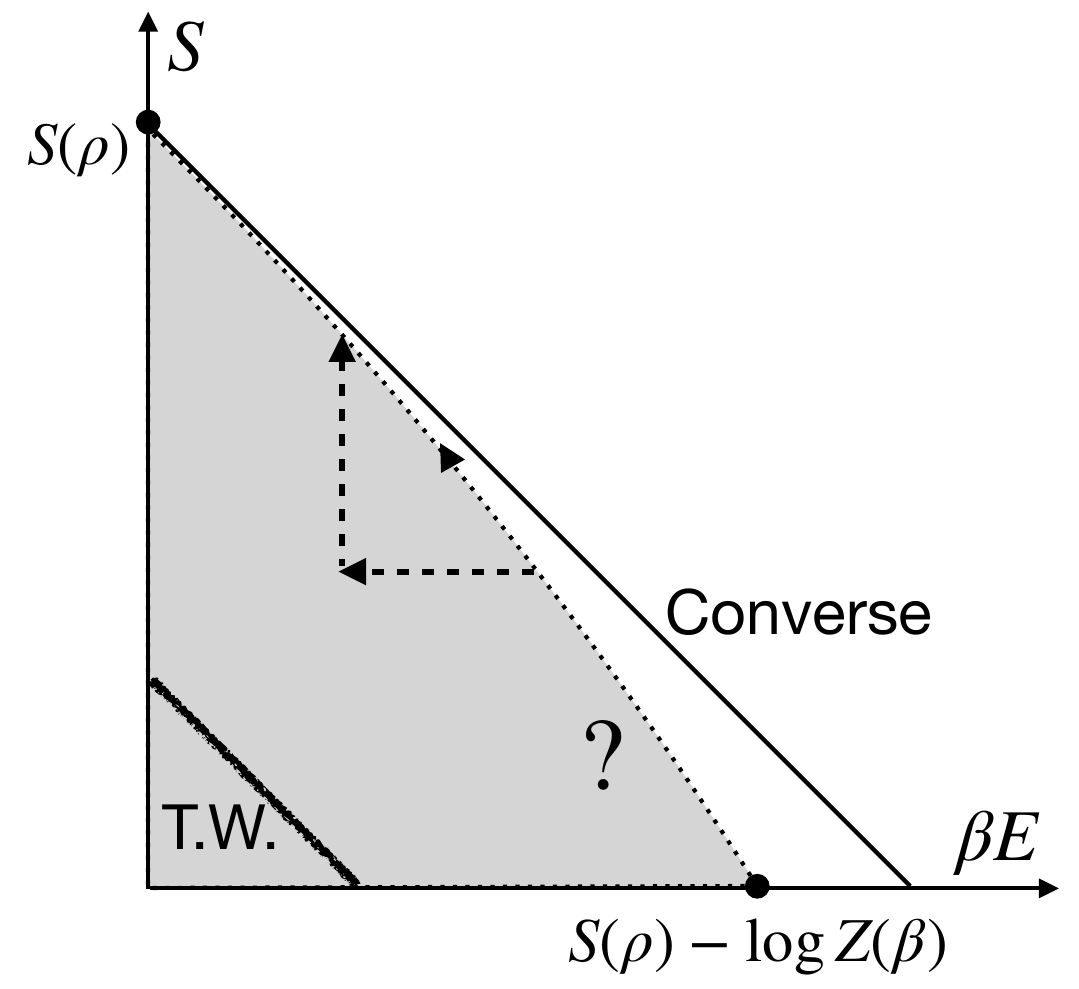}
    \caption{\textbf{Achievable region?} We schematically illustrate in grey the region consisting of all achievable pairs $(\beta E,S)$. The dotted line depicts the optimal protocols. We are certain that $S(\rho)$ is achievable for the optimal QIT protocol and $S(\rho)-\log Z$ is achievable for the optimal QET protocol. The converse bound we prove is plotted as the diagonal solid line. The dashed triangle illustrates that as one changes among the optimal protocols, the gain of teleporting information is compensated by the loss of teleported energy. The perturbative trade-off we found for traversable wormhole teleportations is illustrated as moving on the line in the corner. }
    \label{fig:pareto}
\end{figure}

We leave the proof in Appendix~\ref{app:theorem1}. This trade-off bound is tight. It is saturated by Bell measurements, and meanwhile Bob cannot extract energy. On the other hand, when Alice performs product measurements on $AA'$ and measures the basis $\{\ket{x}_A\}$, Bob can extract the most amount of energy by applying the unitary that maps the steered state $\ket{E_x}_B$ to the ground state, and meanwhile quantum information cannot be teleported. The optimal QET, however, does not saturate the bound, leaving a gap of size $\log Z(\beta)$ (cf. Fig.~\ref{fig:pareto} and discussions later).

We now consider any pure entangled state $\rho_{AB}$ for QET and QIT. Since $\rho_{AB}$ does not have an explicit temperature associated, we need an effective $\beta_*$ to convert between energy and entropy. Importantly, such an effective $\beta_*$ should only depend on the Hamiltonian and the state. We define $\beta_*$ by demanding
\begin{equation}\label{eq:effective_temp}
    S(e^{-\beta_*H}/Z)=S(\rho_B)\ .
\end{equation}
In words, $\beta_*$ is an effective inverse temperature such that the thermal state w.r.t. Bob's Hamiltonian $H$ at temperature $1/\beta_*$ has the same von Neumann entropy as $\rho_B$. When $\rho=\tau_\beta$, we have $\beta_*=\beta$ as desired. 

Operationally, this effective temperature $1/\beta_*$ is relevant in characterizing the maximal unitarily extractable energy over identical and independent distributed (i.i.d.) copies of the state. This quantity (known as the regularized ergotropy~\cite{allahverdyan2004maximal}) is defined as
\begin{equation}
    E^\mathrm{reg}(\rho):=\lim_{n\to\infty}\frac1n \left(\tr\, H\rho^{\otimes n} -\min_U\tr\, HU\rho^{\otimes n}U^\dagger \right)\ .
\end{equation}
It was shown by Alicki and Fannes~\cite{Alicki13} that
\begin{equation}\label{eq:alicki}
    E^\mathrm{reg}(\rho)= \frac{1}{\beta_*}S(\rho||\tau_{\beta_*})\ .
\end{equation}

The fact that $\beta_*$ shows up in the thermodynamic limit suggests that for general entangled states, it is more sensible to study the trade-off for many copies of the resource state $\rho^{\otimes n}_{AB}$ with the total Hamiltonian $\sum_{i=1}^n H^{(i)}$ where each $H^{(i)}$ is identical to Bob's Hamiltonian $H$ but they act on different tensor factors. 

Note that \eqref{eq:alicki} implies even when $\rho$ is passive, Bob could locally extract energy from $\rho^{\otimes n}$ for some large enough $n$ without Alice's help. We thus need to subtract~\eqref{eq:alicki} from Bob's total extracted energy when measuring how much energy is teleported. In the thermodynamic limit, we are thence interested in how the expected teleported energy per copy is in competition with the teleported information per copy. We prove the following result.
\begin{thm}\label{thm:2}
Let $H$ be a bounded Hamiltonian of system $B$.  Given $n$ i.i.d. copies of a pure entangled state $\rho_{AB}^{\otimes n}$ and Bell states $\ket{\Phi}_{A'R}^{\otimes n}$, any teleportation protocol, that consists of any measurement with rank-$1$ projective operators $\{\Pi_x\}$ on $A^n{A'}^n$ followed by any conditional unitaries $\{U_x\}$ on $B^n$, satisfies
\begin{equation}\label{eq:result2}
    S^\mathrm{reg}(B)+\beta_*\Delta E^\mathrm{reg}  \le S(B)_\rho\ ,
\end{equation}
where $\beta_*$ is defined via~\eqref{eq:effective_temp}, $S^\mathrm{reg}(B):=
\sum_x p_x\lim_{n\to\infty}\frac1n S(B^n)_{\Pi_x(\rho^{\otimes n})\Pi_x/p_x}$ is the expected regularized entanglement entropy of the teleported state on $R^nB^n$, and~$\Delta E^\mathrm{reg}:=\lim_{n\to\infty}\frac1n\sum_x p_x E_x - E^\mathrm{reg}(\rho_B)$ is the expected regularized teleported energy. 
\end{thm}

We leave the proof in Appendix~\ref{app:theorem2}. We reiterate that since an $E^\mathrm{reg}(\rho_B)$ amount of energy can be extracted without any teleportation protocol in the asymptotic regime, it is only fair to subtract this contribution from the energy in~\eqref{eq:result2}.  Only copies of the thermal state remain passive for all $n$, and they are called \emph{completely passive}~\cite{lenard1978thermodynamical,pusz1978passive}. For them, the remainder term~\eqref{eq:alicki} vanishes and~\eqref{eq:result2} reduces to the previous result~\eqref{eq:result1}.

\emph{Discussions.}---An operational explanation of the trade-off bounds is as follows. The optimal protocol for QIT cannot work well for QET and vice versa. This is because for QIT, Alice wants a measurement that reveals as little information as possible so as to ``wire up'' the quantum correlation between Bob and the reference as neatly as possible; whereas for QET, Alice wants a measurement that reveals some information that pertains to the Hamiltonian, to instruct Bob to extract energy accordingly. These goals contradict each other so one has to compromise when limited entanglement resources are available. Therefore, the total entanglement shared among Alice and Bob determines the optimal overall performance of both QET and QIT.

We also propose a more physical explanation inspired by the perturbative trade-off observed in the traversable wormhole protocol (cf. Fig.~\ref{fig:pareto}). We can rearrange \eqref{eq:result1} to $S(B)\le\beta (E_{\max}-E-F)\ $, where $E_{\max}=\tr\,H\tau_\beta$ is the maximal extractable energy from Bob's system (assuming the ground state energy is zero), and $F=-\log Z<0$ is Bob's equilibrium free energy at $\beta$. The same manipulation also applies for~\eqref{eq:result2}. This bound on entropy can be understood as a Bekenstein bound, that the energy cost of sending $S(B)$ amount of information is the energy deficit $(E_{\max}-E)$ (up to the additive constant of $F$). Bob could have gained had they used the optimal QET protocol. It always consumes positive energy to transmit information, which acts against QET. The same principle is manifest in the bulk picture of the traversable wormhole protocol as we have shown above.

%Achievability
What we manage to prove is a converse bound to the overall performance of QET+QIT measured by the pair $(\beta E, S)$ (or the regularized version $(\beta_* \Delta E^\mathrm{reg}, S^\mathrm{reg})$). It would be interesting to find the achievable region in the $\beta E-S$ plane spanned by all viable protocols, or deduce its general properties. We schematically illustrate the achievable region in Fig.~\ref{fig:pareto}. We know that for QET the bound is not achievable with a gap of $F$, so one could hope to tighten the converse with different techniques. 

In the traversable wormhole teleportation, neither QET nor QIT gets close to the optimal bound that is of order $\O(N)$. Instead, their performances are controlled by the coupling strength $g$ as shown in~\cite{maldacena2017diving} for QIT and in~\eqref{Energychange} for QET. This is because the parameters regime is chosen such that the protocol admits a simple semiclassical bulk description. It would be nice to push the protocol beyond the semiclassical regime towards saturating the bound. 

%More general QET settings.
Inspired by the traversable wormhole teleportation, we only studied non-interacting Hamiltonians. It is interesting to extend the trade-off bound to the case of interacting Hamiltonian where QET and QIT can be implemented on, for instance, the ground state\footnote{This is the conventional setting for QET~\cite{hotta2008protocol,hotta2008quantum}. Our setup with an excited entangled resource state was also studied by Hotta in~\cite{hotta2014quantum}.}. One technical difficulty is to assign an effective temperature that depends on the Hamiltonian, which is indispensable for converting between energy and entropy. Our proposal~\eqref{eq:effective_temp} could not work for the lack of the local Hamiltonian. 

One could in principle also look for different formulations of the trade-off that is not of our thermodynamic form. Interestingly, a concurrent work by Adam Brown also demonstrates the competition between information and energy transmission along a relativistic string~\cite{brown2024channel}.  We conjecture that the fight between them holds universally for any joint QET+QIT protocol.

\emph{Acknowledgments.}---We are grateful to Zhenbin Yang and Douglas Stanford for very helpful comments and discussions on traversable wormholes. We also thank Adam Brown, Raphael Bousso, Patrick Hayden, Henry Lin, Jingru Lu, and Jianming Zheng for the discussions. JW would like to thank the support from AFOSR (award FA9550-19-1-0369).

\bibliographystyle{unsrtnat}
\bibliography{evs}

\appendix

\widetext

\section{Double trace Deformation as LOCC}\label{app:LOCC}
In this section, we show that the double trace deformation $e^{ig\frac1K\sum_{i=1}^K O_i^AO_i^B}$ has the same effect on Bob as an LOCC operation consisting of sequential measurements made by Alice. More precisely, we mean that we can find an LOCC scheme, in which Alice performs some measurements and sends the result to Bob, and then Bob will apply some conditional unitary, such that it has the same effect on Bob's reduced state as the double-trace-deformed reduced state. Note that their effects on Alice's side are certainly different, but we only care about the energy and the entropy of Bob in the end of the protocol. 

In particular, we demonstrate the case where the underlying boundary quantum mechanical system is the SYK model consisting of Majorana fermions, and we choose $\{O_i^AO_i^B\}_i = \{\psi_i^A\psi_i^B\}_i$ to be simple fermion operators. 

The first problem is that $\psi_x$'s are fermions, but it is easier to do local operations with bosons. We can introduce a pair of auxiliary Majorana fermions $\psi_0^A$ and $\psi_0^B$ to make composite bosons. We prepare them in the state $\ket{\text{EPR}_0}$ that satisfies $(\psi_0^A + i \psi_0^B) \ket{\text{EPR}_0} = 0$. Now we can rewrite the double trace deformation acting on the thermal field double state as
\begin{equation}
    e^{g \frac{1}{K} \sum_{i=1}^K \psi_i^A \psi_i^B \psi_0^A \psi_0^B} = \prod_{i=1}^N P^A_{i,\pm} \otimes e^{\mp i g \psi_0^B \psi_i^B}
\end{equation}
acting on $\ket{\tau_\beta}_{AB} \otimes \ket{\text{EPR}_0}$. Here, $P^A_{i,\pm}$ denotes the projection onto state $\ket{O_{i}}$ satisfying $(\psi_i^A \pm i\psi_0^A)\ket{O_{i}}=0$. Thus, the double-trace deformation has the same effect on Bob as an LOCC operation with sequential measurements $\Pi_x:=\Pi_i P_{i,\pm}^A$ made by Alice, with outcomes $x\in\{\pm\}^{\times N}$ and corresponding conditional unitary $U_x:=\Pi_i e^{\mp i g \psi_0^B \psi_i^B}$ implemented by Bob. 

\section{Traversable wormhole as QET}\label{app:TW}
In this section, we present the details for computing the energy change in Bob's system, using the Schwarzian description of JT gravity~\cite{maldacena2016conformal,KitaevTalks,maldacena2016remarks}. By the arguments presented in the previous section, the teleported energy extracted via LOCC can be calculated by comparing the \emph{energy change} before and after the double trace deformation.

Due to the topological nature of JT gravity, the gravitational degree of freedom is described by a boundary reparametrization mode denoted by $\tau(u)$. We will compute the energy change by first computing in Euclidean (imaginary time) configuration and then analytically continue the result to Lorentzian (real) time. On the Euclidean disk, the saddle point solution is given by $\tau(u)=u$. We expand the Schwarzian variable to $\tau(u)=u+\epsilon(u)$, at leading order, the Schwarzian action becomes
\begin{equation}
    S_E=\frac{C}{2}\int_0^{2\pi} d u \, \epsilon''(u)^2-\epsilon'(u)^2\ .
\end{equation}
Here $C$ is the coupling constant, and we set $\beta=2\pi$ for notational clarity in this computation. To compute the energy change, we want to compute the classical solution with a double trace deformation $e^{ig\frac1K\sum_{i=1}^K \psi_i^A(u_1)\psi_i^B(u_2)}$. In Schwarzian formalism, it can be rewritten as $e^{igG(u_1,u_2)}$, where
\begin{equation}
    G(u_1,u_2)=\left(\frac{\tau'(u_2)\tau'(u_1)}{(\tau(u_2)-\tau(u_1))^2}\right)^\Delta= \Delta\left( \epsilon'(u_2)+\epsilon'(u_1)+(\epsilon(u_1)-\epsilon(u_2))\cot(\frac{u_2-u_1}{2}) \right)\frac{1}{\sin^{2\Delta}(\frac{u_2-u_1}{2})}\ .
\end{equation}

For the energy teleportation protocol that we considered, we need to solve the saddle point of total action $-S_E+igG(0,\tau)$. We leave $\tau$ general here and we will analytically continue it to $\pi +it$ later. The equation of motion for the whole action is
\begin{equation}\label{append:EOM}
    C(\epsilon''''(u)+\epsilon''(u))+ig\Delta\frac{1}{\sin^{2\Delta}(\tau/2)}(\delta'(0)+\delta'(\tau)-(\delta(0)-\delta(\tau))\cot(\tau/2))=0\ .
\end{equation}
We can solve the classical solution explicitly, but for our purpose, we care about how energy
\begin{equation}
    H=C\left[\frac{1}{2}+(\epsilon'(u)+\epsilon'''(u)) \right]
\end{equation}
changes before and after the operator insertion. The only contribution to the energy change before and after operator insertion at $\tau$ comes from the delta function in \eqref{append:EOM}, thus we conclude that the energy change is 
\begin{equation}
    -i g\Delta \frac{\cot(\tau/2)}{\sin^{2\Delta}(\tau/2)}\ .
\end{equation}
Now we can analytically continue to Lorentzian time $\tau_1=\pi +i t$ to obtain the averaged teleported energy $E$ as the negative of the energy change 
\begin{equation}
    E= g\Delta \frac{\tanh(t/2)}{\cosh^{2\Delta}(t/2)}\ .
\end{equation}
Putting back the $\beta$ dependence yields
\begin{equation}
    E= g\Delta\left(\frac{2\pi}{\beta}\right)^{2\Delta+1} \frac{\tanh(\frac{\pi}{\beta}t)}{\cosh^{2\Delta}(\frac{\pi}{\beta}t)}\ .
\end{equation}

In conclusion, Bob can extract energy if he applies the unitary on his share of the thermal field double state with a delay $t>0$.

\section{Proof of Theorem 1}\label{app:theorem1}

\begin{proof}
We can rewrite the thermal entangled state as $\sqrt{\tau_\beta}_B\ket{\id}_{AB}$ where the square root of the density matrix $\tau_\beta=\sum_x e^{-\beta H} \ketbra{E_x}{E_x}_B$ acts on the (super-normalized) maximally entangled state $\ket{\id}_{AB}=\sum_x\ket{x}_A\ket{x}_B$. In this way, it is manifest that Bob's density matrix is $\tau_\beta$. 

Any rank-one projector $\Pi_x$ on $AA'$ can be written as $\Pi_x=\ketbra{\Pi_x}{\Pi_x}_{AA'}$. By virtue of the Schmidt decomposition of $\ket{\Pi_x}_{AA'}$, we can write any pure state as $\ket{\Pi_x}_{AA'}=\sqrt{\sigma_{x\,A}}\otimes U_{x\,A'}\ket{\id}_{AA'}$ for some positive operator $\sigma_x\ge 0$ on $A$ and some unitary $U_x$ on $A'$. Note that $\{\sigma_x\}$ forms a POVM set on $A$, because
$$\id_A=\tr_{A'}\sum_x \Pi_x= \sum_x \tr_{A'} \left( \sqrt{\sigma_{x\,A}}\otimes U_{x\,A'}\ketbra{\id}{\id}_{AA'}\sqrt{\sigma_{x\,A}}\otimes U^\dagger_{x\,A'}\right) = \sum_x \sigma_{x\,A}\ .$$
The probability of Alice measuring $\Pi_x$ is
\begin{equation}
    p_x= \bra{\tau_\beta}_{AB}\bra{\Phi}_{A'R}\sqrt{\sigma_{x\,A}}\otimes U_{x\,A'}\ketbra{\id}{\id}_{AA'}\sqrt{\sigma_{x\,A}}\otimes U^\dagger_{x\,A'}\ket{\tau_\beta}_{AB}\ket{\Phi}_{A'R}=\tr\, \sigma_x^\top\tau_\beta=\tr\, \sigma_x\tau_\beta\ , \ 
\end{equation}
where we use the fact that $\tau_\beta^\top=\tau_\beta$ and $\tr\, \sigma_x^\top\tau_\beta^\top=\tr\, \sigma_x\tau_\beta$ in the last step.

The post-measurement state reads,
\begin{equation}
    \ket{\rho_x}_{BR}:=  \frac{1}{\sqrt{p_x}}\bra{\id}_{AA'}\sqrt{\sigma_{x\,A}}\otimes U^\dagger_{x\,A'}\ket{\tau_\beta}_{AB}\ket{\Phi}_{A'R}= \frac{1}{\sqrt{p_x}}(\sqrt{\sigma_{x}}^\top\sqrt{\tau_\beta})_B\otimes\bar U_{x\,R}\ket{\id}_{BR}\ ,
\end{equation}
and the post-measurement marginal on $B$ reads,
\begin{equation}\label{eq:postmmt}
    \rho_x:=\tr_R\,\rho_{BR|x}=\sqrt{\tau_\beta}\sigma_x^\top\sqrt{\tau_\beta}/p_x \ ,
\end{equation}
and they average to 
\begin{equation}\label{eq:sumtotau}
    \sum_x p_x\rho_x = \sum_x\sqrt{\tau_\beta}\sigma_x^\top\sqrt{\tau_\beta} = \tau_\beta\ ,
\end{equation}
as $\{\sigma_x\}$ forms a POVM set on $A$ and so does $\{\sigma_x^\top\}$ on $B$.

Consider now Bob's extracted energy upon learning Alice's measurement outcome $x$,
\begin{equation}\label{eq:energyextract}
  E_x=\tr\,H(\rho_x -U_x\rho_xU_x^\dagger) = \frac1\beta\left( S(\rho_x||\tau_\beta) - S(U_x\rho_xU_x^\dagger||\tau_\beta)\right)\le \frac1\beta S(\rho_x||\tau_\beta) \ ,
\end{equation}
where $S(\rho_x||\tau_\beta):=\beta \langle H\rangle_{\rho_x}+\log Z-S(\rho_x)$, and we have arranged the energy extracted as the difference between two relative entropies so we can dump the second relative entropy term which is always positive. We therefore have
\begin{equation}
    S(\rho_x)+\beta E_x\le -\langle\log\tau_\beta\rangle_{\rho_x}\ .
\end{equation}
Taking the expectation over the measurement outcome,
\begin{equation}
    \sum_x p_x \left[S(\rho_x)+\beta E_x\right]\le -\sum_x p_x \langle\log\tau_\beta\rangle_{\rho_x} = -\langle\log\tau_\beta\rangle_{\tau_\beta} = S(\tau_\beta)\ ,
\end{equation}
where we have used \eqref{eq:sumtotau}.
\end{proof}

\section{Proof of Theorem 2}\label{app:theorem2}

For the proof, we will be using an important quantity called the max-relative entropy~\cite{renner2008security,datta2009min},
\begin{equation}
    S_{\max} (\rho||\sigma):= \log\min\{\alpha: \rho \le \alpha\,\sigma\} \ .
\end{equation}
The definition essentially says that 
\begin{equation}\label{eq:op_inequality}
    \rho \le 2^{S_{\max} (\rho||\sigma)}\,\sigma \ 
\end{equation}
is the tightest possible operator inequality upper-bounding $\rho$ with $\sigma$. Often it is more operational to use the smoothed max-relative entropy,
\begin{equation}
    S_{\max}^\eps (\rho||\sigma):=\min_{\rho'\approx^\eps\rho}S_{\max} (\rho'||\sigma) \ ,
\end{equation}
where $\rho'$ is close to $\rho$ in the purified distance $P(\rho',\rho):=\sqrt{1-F(\rho',\rho)}$.

The intuition behind working with the max-relative entropy is that: By taking the logarithm of the operator inequality~\eqref{eq:op_inequality} and then taking the expectation value w.r.t. $\rho$, we already get something close to what we want. To relate the max-relative entropy to the standard Umegaki relative entropy, we need the following typicality property.
\begin{lem}[Fully quantum asymptotic equipartition~\cite{tomamichel2009fully,tomamichel2015quantum}]\label{lem:aep}
    Let $0<\eps<1$, and $\rho,\sigma$ be any quantum states such that $\mathrm{supp}(\rho)\subseteq\mathrm{supp}(\sigma)$, then
    \begin{equation}
        \lim_{n\to\infty}\frac1n S_{\max}^\eps (\rho^{\otimes n}||\sigma^{\otimes n})=S(\rho||\sigma)\ .
    \end{equation}
\end{lem}

Some other lemmas shall prove handy as well.

\begin{lem}[Continuity of energy]\label{lem:energy}
Let $H$ be a bounded Hamiltonian with ground state energy set at zero, and $\rho,\sigma$ be two quantum states,
\begin{equation}
    |\tr\, H(\rho-\sigma)|\le \frac12||\rho-\sigma||_1||H||_\infty\le P(\rho,\sigma)||H||_\infty\ .
\end{equation}
\end{lem}
This is a direct consequence of H\"older's inequality and that the purified distance is larger than the trace distance.

\begin{lem}[Continuity of entropy~\cite{fannes1973continuity,audenaert2007sharp}]\label{lem:entropy}
Let $\rho,\sigma$ be two $d$-dimensional quantum states such that their trace distance is close, $\frac12||\rho-\sigma||_1 \le\eps$, then their von Neumann entropy is also close,
\begin{equation}
    |S(\rho)-S(\sigma)|\le \eps\log(d-1)+ h_2(\eps)\ .
\end{equation}
where $h_2(\eps):=-\eps\log\eps-(1-\eps)\log(1-\eps)$ is the binary entropy function.
\end{lem}

Now we are ready to prove Theorem 2.
\begin{proof}
%Without loss of generality, we can assume $\rho_{AB}$ to be pure. Because given any mixed state $\rho_{AB}$, we can consider the purification $\rho_{ABC}$. Then any measurement on $A$ is also a measurement on $AC$ that acts trivially on $C$. It is therefore sufficient to consider pure resource states as long as we allow all measurement instruments. 

Just like in the last proof, the Schmidt decomposition guarantees that any bipartite pure state can be written as
\begin{equation}
    \ket{\rho_{AB}} = U_A\otimes\sqrt{\rho_B}\ket{\mathbb{I}}_{AB}\ .
\end{equation}

Let us consider an arbitrary QET protocol $(\Pi_x,U_x)$, where $\{\Pi_x\}$ are rank-one projectors on $A^nA'^n$. Following the same argument as the last proof, we can show that upon obtaining outcome $x$, the post-measurement state on $B^n$ reads
\begin{equation}
    \rho_{B^n|x}:=\sqrt{\rho_B^{\otimes n}}\sigma_x\sqrt{\rho_B^{\otimes n}}/p_x \ ,\quad p_x:=\tr_{AB}\,\Pi_x\rho_{AB}^{\otimes n} \Pi_x=\tr_B\, \sigma_x\rho_B^{\otimes n}\ , \quad\sigma_x:=\tr_A U^\top \Pi_x\bar U\ ,
\end{equation}
where $\{\sigma_x\}$ forms a POVM set on $B^n$.
The ensemble $\{p_x,\rho_{B^n|x}\}$ averages to $\rho_B^{\otimes n}$, 
\begin{equation}\label{eq:ensemble_average}
    \sum_x p_x\rho_{B^n|x}=\rho_B^{\otimes n}\ .
\end{equation}

Consider the smooth max-relative entropy between $\rho_B^{\otimes n}$ and $\tau^{\otimes n}_\beta:=e^{-\beta\sum_{i=1}^n H^{(i)}}/Z$ for any $\beta$, where $Z$ is the partition function $Z=\tr e^{-\beta\sum_{i=1}^n H^{(i)}}$ throughout this proof. Let $\rho'_{B^n}$ be the minimizer of it,
\begin{equation}
    S_{\max}^\eps(\rho_B^{\otimes n}||\tau^{\otimes n}_\beta) = S_{\max}(\rho'_{B^n}||\tau^{\otimes n}_\beta)\ .
\end{equation}
By definition, we have $\frac12||\rho'_{B^n}-\rho_B^{\otimes n}||_1\le P(\rho'_{B^n},\rho_B^{\otimes n})\le\eps$, and
\begin{equation}
    \rho'_{B^n}\le 2^{S^\eps_{\max}(\rho_B^{\otimes n}||\tau^{\otimes n}_\beta)}\tau^{\otimes n}_\beta\ .
\end{equation}
Taking the logarithm gives,
\begin{equation}
    \log\rho'_{B^n}\le S^\eps_{\max}(\rho_B^{\otimes n}||\tau^{\otimes n}_\beta)-\log Z-\beta\sum_{i=1}^n H^{(i)}\ .
\end{equation}
Then taking the expectation value on the state $\rho'_{B^n}$ yields
\begin{equation}
    \langle\log\rho'_{B^n}\rangle_{\rho'_{B^n}}=-S(\rho'_{B^n}) \le S^\eps_{\max}(\rho_B^{\otimes n}||\tau^{\otimes n}_\beta)-\log Z-\beta \langle \sum_{i=1}^n H^{(i)}\rangle_{\rho'_{B^n}}\ .
\end{equation}
Since $\rho'_{B^n}$ and $\rho_B^{\otimes n}$ are close in trace distance, we can bound the energy difference between them by Lemma~\ref{lem:energy},
\begin{equation}
    \langle \sum_{i=1}^n H^{(i)}\rangle_{\rho'_{B^n}}\ge \langle \sum_{i=1}^n H^{(i)}\rangle_{\rho_B^{\otimes n}} - n\eps||H||_\infty\ ,
\end{equation}
where $||H||_\infty$ is finite by assumption. It follows that 
\begin{equation}
    -S(\rho'_{B^n}) \le S^\eps_{\max}(\rho_B^{\otimes n}||\tau^{\otimes n}_\beta)-\beta \langle \sum_{i=1}^n H^{(i)}\rangle_{\rho_B^{\otimes n}}-\log Z+n\eps||H||_\infty\ .
\end{equation}
Using~\eqref{eq:ensemble_average} and the linearity of the energy expectation, we have
\begin{equation}
    -S(\rho'_{B^n}) \le S^\eps_{\max}(\rho_B^{\otimes n}||\tau^{\otimes n}_\beta)-\beta \sum_x p_x\langle \sum_{i=1}^n H^{(i)}\rangle_{\rho_{B^n|x}}-\log Z+n\eps||H||_\infty\ .
\end{equation}
Using the definition of extracted energy $E$, we have
\begin{equation}\label{eq:step1}
    -S(\rho'_{B^n}) \le S^\eps_{\max}(\rho_B^{\otimes n}||\tau^{\otimes n}_\beta) -\beta\sum_x p_x E(\rho_{B^n|x})-\beta\sum_x p_x\langle\sum_{i=1}^n H^{(i)}\rangle_{U_x\rho_{B^n|x}U_x^\dagger} -\log Z+n\eps||H||_\infty\ ,
\end{equation}
where $U_x$ is the optimal energy extractor for $\rho_{B^n|x}$.

Consider the relative entropy $S(U_x\rho_xU_x^\dagger||\tau_\beta)$. Its positivity implies that
\begin{equation}
    S(U_x\rho_{B^n|x}U_x^\dagger||\tau_\beta) = \beta\langle \sum_{i=1}^n H^{(i)}\rangle_{U_x\rho_{B^n|x}U_x^\dagger}+\log Z-S(\rho_{B^n|x})\ge 0 \ .
\end{equation}
Plugging the above inequality to \eqref{eq:step1} yields
\begin{equation}
    -S(\rho'_{B^n}) \le S^\eps_{\max}(\rho_B^{\otimes n}||\tau^{\otimes n}_\beta)-\beta\sum_x p_x E(\rho_{B^n|x})-\sum_x p_xS(\rho_{B^n|x})+n\eps||H||_\infty\ .
\end{equation}
Using Lemma~\ref{lem:entropy}, we can change the entropy $S(\rho'_{B^n})$ on the LHS to $S(\rho_B^{\otimes n}) (= nS(\rho_B))$, because their trace distance is smaller than $\eps$.
\begin{equation}
    -nS(\rho_B)-\eps n\log d-h_2(\eps)\le S^\eps_{\max}(\rho_B^{\otimes n}||\tau^{\otimes n}_\beta)-\beta\sum_x p_xE(\rho_{B^n|x})-\sum_x p_xS(\rho_{B^n|x})+n\eps||H||_\infty\ .
\end{equation}

Now taking the per-copy average and sending $n\to\infty$ yield,
\begin{equation}
   S(B)_\rho- \lim_{n\to\infty}\frac{\beta}{n}\sum_x p_xE(\rho_{B^n|x})+\lim_{n\to\infty}\frac1nS^\eps_{\max}(\rho_B^{\otimes n}||\tau^{\otimes n}_\beta) \ge \lim_{n\to\infty}\frac1n \sum_x p_xS(\rho_{B^n|x})-\eps(\log d+||H||_\infty)\ .
\end{equation}
Then applying Lemma~\ref{lem:aep} yields
\begin{equation}
    S(B)_\rho \ge \lim_{n\to\infty}\frac{\beta}{n}\sum_x p_xE(\rho_{B^n|x})-S(\rho_B||\tau_{\beta})+\lim_{n\to\infty}\frac1n \sum_x p_xS(\rho_{B^n|x})-\eps(\log d+||H||_\infty)\ .
\end{equation}
which according to our definition is,
\begin{equation}
    S(B)_\rho \ge \beta\left(\frac{1}{n}\sum_x p_xE(\rho_{B^n|x})-\frac{1}{\beta}S(\rho_B||\tau_{\beta})\right)+S^\mathrm{reg}(B)-\eps(\log d+||H||_\infty)\ .
\end{equation}

This inequality holds for all $\eps>0$. We can take the infimum over $\eps$ to get rid of the last remainder term on the RHS. Also, this inequality holds for all $\beta>0$. We choose $\beta=\beta_*$, such that $\frac{1}{\beta_*}S(\rho_B||\tau_{\beta_*})$ acquires the operational meaning of being extractable energy without Alice's message. We have thus shown the trade-off bound (8). 
\end{proof}

\section{Teleportation schemes with general measurements.}\label{app:mmts}
The proofs we give above consider only the optimal rank-one projective measurements. When the measurement operators $\{\Pi_x\}$ are not rank-one projectors, the post-measurement state $\rho_{RB|x}$ is mixed. 

Consider any measurement instruments described by a set of Kraus operators $\{\Pi_x\}$. Let the corresponding POVM element on $AA'$ be $\Sigma_x:=\Pi_x^\dagger \Pi_x$. Let $\sigma_x=\tr_{A'}\Sigma_x$, which form a POVM set on~$A$. Upon obtaining the outcome $x$, the post-measurement state reads
\begin{equation}
    \rho_{RA'AB|x}= \Pi_x\ket{\Phi}_{RA'}\ket{\tau_\beta}_{AB}/p_x,\quad p_x=\bra{\Phi}_{RA'}\bra{\tau_\beta}_{AB}\Sigma_x\ket{\Phi}_{RA'}\ket{\tau_\beta}_{AB}=\tr\,\tau_\beta\sigma_x\ ,
\end{equation}
and the post-measurement marginal on B reads,
\begin{equation}
    \rho_x:=\sqrt{\tau_\beta}\sigma_x^\top\sqrt{\tau_\beta}/p_x \ ,\quad p_x=\tr\,\tau_\beta\sigma_x\ ,
\end{equation}
which is identical to \eqref{eq:postmmt}. By the same steps as in the proof for Theorem 1, it follows that 
\begin{equation}
    \sum_x p_x \left[S(B)_{\rho_x}+\beta E_x \right] \le S(B)_{\tau_\beta}\ .
\end{equation}
Hence, the claims we made in Theorem 1 and Theorem 2 still hold for general measurements.

The caveat is that the entropy $S(B)_{\rho_x}$ doesn't directly measure the entanglement for mixed states. Nonetheless, it upper-bounds the entanglement of formation of $\rho_{AB|x}$, so we have
\begin{equation}
    \sum_x p_x \left[E_f(\rho_x)+\beta E_x \right] \le S(B)_{\tau_\beta}\ .
\end{equation}

\section{Variants of the trade-off relation (4)}\label{app:variant}

There are some variants of our trade-off relation (4). Let $\rho_B:=\sum_x p_x \rho_x$ be the average output state of the teleportation protocol. Then we have
\begin{equation}\label{eq:result1.5}
  E_C(\rho)+\beta E\le E_f(\rho) +\beta E \le S(B)_{\tau_\beta}\ ,
\end{equation}
where $E_f(\rho):=\min_{\{p_i,\rho_i\}}\sum_i p_i S(\rho_i)$ is the entanglement of formation~\cite{hill1997entanglement,Wootters1998,horodecki2009quantum}. The bound in the middle follows from (4) because $S_f(\rho)$ is defined as the minimization over all ensemble decompositions $\{p_i,\rho_i\}$ of $\rho$, and $\{p_x,\rho_x\}$ is one such instance. Entanglement of formation measures how much entanglement (measured in the number of qubit Bell pairs) is needed on average to prepare an entangled state $\rho$. It asymptotes to the entanglement cost~\cite{bennett1996mixed,hayden2001asymptotic} $E_C(\rho)=\lim_{n\to\infty} E_f(\rho^{\otimes n})/n$. Since $E_f$ is subadditive, we can further lower-bound it by $E_C(\rho)$, which yields the bound on the left.

\end{document}